\documentclass[10pt,preprint2]{aastex}%
\begin{document}

\title{Q0906+6930: The Highest-Redshift Blazar}
\author{Roger W. Romani, David Sowards-Emmerd, Lincoln Greenhill\altaffilmark{1} \& Peter Michelson}
\affil{Department of Physics, Stanford University, Stanford, CA 94305-4060}
\altaffiltext{1}{Permanent Address: Harvard-Smithsonian Center for Astrophysics,
60 Garden St, Cambridge, MA 02138}
\email{rwr@astro.stanford.edu, dse@darkmatter.stanford.edu,lincoln@cfa.harvard.edu,peterm@stanford.edu}

\begin{abstract}
	We report the discovery of a radio-loud flat-spectrum QSO
at z=5.47 with properties similar to those of the {\it EGRET} $\gamma$-ray blazars. 
This source is the brightest radio QSO at z$>$5, with a pc-scale radio jet and 
a black hole mass estimate $\ga 10^{10}M_\odot$.  It appears to be the 
most distant blazar discovered to date. High energy observations of this
source can provide powerful probes of the background radiation in the early universe.
\end{abstract}


\section{Introduction}

	AGN classification is still somewhat heuristic, but in
the unified model, blazars are believed to be sources viewed close to the
axis of a powerful relativistic jet \citep[][and references therein]{up95}.
They thus have compact
flat spectrum radio counterparts, with apparent superluminal motion at VLBI 
scales. Optically, the sources are often variable, exhibit significant 
polarization and show either broad emission lines (flat spectrum radio quasar) 
or continuum-dominated (BL Lac) spectra. Perhaps
the most interesting consequence of the jet/line-of-sight alignment is the
observability of a Compton scattered component at X-ray to GeV or
(for nearby sources) TeV energies. The interaction of the high energy
jet particles and radiation with surrounding photon fields allows unique
probes of the extragalactic background light (EBL).

	One of the principal discoveries of the {\it EGRET} experiment
on the {\it Compton Gamma Ray Observatory} was the large population of 
blazars emitting at GeV energies. These represented the largest identified
population in the third {\it EGRET} (3EG) catalog; still many high latitude
sources remained unidentified.  We have developed a method to extend 
identifications to fainter radio flux levels \citep{srm03,srmu04}, finding 
counterparts for 70\% of the $|b|>10^\circ$ sources. Follow-on observations 
show that most of these are previously unidentified blazars.
Upcoming GeV $\gamma$-ray missions, especially {\it GLAST}, 
will have much higher sensitivity and are expected to detect several thousand 
blazars.  This substantially exceeds the total number of blazars cataloged to date
\citep{lan01}. Thus in preparation for this mission, we have selected radio
(\& X-ray) sources with properties like the {\it EGRET}\, blazars and
are obtaining optical identifications and spectroscopy \citep[][SRM04]{srm04}.

	To prioritize the optical observations, we have also developed
a method for estimating the probability that a given position has an excess
of $\gamma$-ray photons in the {\it EGRET} survey observations, including the 
effect of strong variability.  The radio blazar candidates selected above show, as a
set, a clear excess of $\gamma$-ray flux over random sky positions (SRM04);
individually, they are well below the 4$\sigma$ criterion for 
inclusion in the 3EG catalog. For example, radio blazar candidates with $\ge$75\% 
probability of 
{\it EGRET} $\gamma$-ray excess are found at $\sim 0.012/\circ^2$, and thus may
represent 10-20\% of the new sources detectable to {\it GLAST}. These are
prime candidates for optical follow-up; we find that most are
indeed flat spectrum radio quasars with redshifts 1-2.5. Q0609+6930,
however, has a very high redshift of 5.47. Observations described below support
a blazar identification for this source; this would be the highest
redshift blazar discovered to date.

\section{The discovery of Q0906+6930}

In selecting $\gamma$-ray blazar candidates
we start from compact 8.4GHz sources (from CLASS, Meyers et al. 2003 or 
from our own VLA snapshots) and identify flat/inverted spectrum 
sources using NVSS counterparts. Our selection algorithm also applies 
a weighting toward X-ray sources sources detected in the {\it ROSAT}
All-Sky Survey (RASS); this is weak as many of the {\it EGRET}
blazars are below the RASS survey threshold. This identification scheme differs
from other blazar surveys \citep[eg. DXRBS,][]{lan01} and gives a much
higher correlation with the 3EG sources. We follow up with optical 
spectroscopy obtained with the Marcario Low Resolution Spectrograph 
\citep[LRS;][]{hill98} at the Hobby-Eberly telescope \citep[HET;][]
{ram98}, obtaining source classifications and redshifts.

\begin{figure}
\plotone{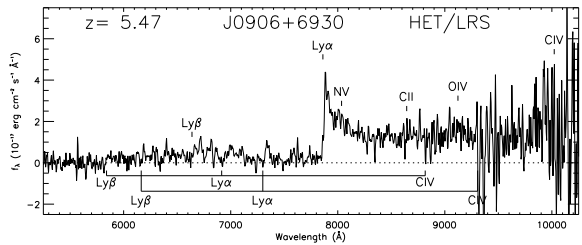}
\figcaption{\small
HET/LRS spectrum of Q0906+6930.
}
\end{figure}

	Q0906+6930 has a strongly inverted spectrum $\alpha_{1.4/8.4}=-0.4$
$(S_\nu \propto \nu^{-\alpha})$
with a CLASS-epoch flux of $S_{8.4} = 190$mJy. The radio position is
09:06:30.75 +69:30:30.8 (J2000.0).  This source was cataloged as
compact at the $\sim 0.1^{\prime\prime}$ scale, but the short CLASS
snapshots provide only limited $\sim 1$mJy sensitivity to flux on arcsecond
scales. This source is just included in our `sub-threshold' $\gamma$-ray
target list, with a likelihood analysis of the $\gamma$-ray counts giving
a 75\% probability for emission in excess of background at this position.

\subsection{HET/LRS optical spectrum}

	We obtained 2$\times$300s exposure with the HET LRS on 1/18/04,
under poor $\sim 2^{\prime\prime}$ conditions, which nonetheless showed 
a classic Ly$\alpha$ forest-absorbed QSO spectrum with high $z\approx 5.5$ 
redshift. Follow-on exposures of 2$\times$600s on 1/27/04 with a 300\,l/mm 
grating, $2^{\prime\prime}$ slit and 5150\AA~ long-pass filter (to preclude
2nd order contamination) and improved $\sim 1^{\prime\prime}$ seeing
produced the spectrum shown in figure 1. These data were subject to optimal 
extraction, calibration and telluric correction using standard IRAF routines. 
Uncorrectable fringing and imperfect sky subtraction introduce large noise 
beyond 9500\AA.  The resulting spectrum has a dispersion of 4\AA\,/pixel and 
an effective resolution of 16\AA\,, covering $\lambda\lambda$5200-10,000\AA.

In the spectrum Ly$\alpha$ is strong, but absorbed; we estimate a pre-absorption
rest equivalent width of 50-60\AA. The redshift, z=5.47$\pm0.02$,
is best constrained by the NV and OIV lines; the OIV rest EW is 22\AA\, and 
the kinematic width is $W_{FWHM} = 5,000\pm 500$km/s. The UV continuum luminosity 
at 1350\AA\, is $\lambda {\rm L_{1350}} \approx 5 \times 10^{47}$erg/s
(we use $H_0=71$km/s/Mpc, $\Omega_m=0.27$, $\Omega_\Lambda=0.73$).
CIV is not clearly detected, so if we use the OIV
kinematic width in the \citet{ves02} UV M-L relationship we obtain
${\rm Log}M = 6.2 + {\rm Log}[({\rm W_{FWHM}/10^3 km/s})^2(\lambda L_{1350}/10^{44}
{\rm erg/s})^{0.7}] \approx 10.2$. This is near the upper end of masses inferred
for high z sources and formation of such a high $M$ black hole
after $\sim 1$Gyr is difficult to understand. There are several candidate
CIV absorption line systems redward of Ly$\alpha$. Two candidate systems with
apparent damped Ly lines are marked.  The presence of intervening absorption
raises the possibility that Q0906+6930 is lensed, which could inflate our estimate
of the BH mass.  However, CLASS finds no radio lens, there is no extended emission
visible on the POSS and the source appears point-like in a short pre-spectrum 
acquisition image. Improved optical/near IR spectroscopy is needed to tighten up the
mass estimate, and to improve the identification of the absorption systems as
strong CIV doublets.

\subsection{VLBA Observations and Source SED}

	To search for evidence of compact jet structure that would support
the blazar designation of Q0906+6930, we obtained snapshots with the VLBA
at 2cm and 7mm wavelengths (Figure 2). With 10 stations recording two 8-MHz 
bandpasses at 15.36\,GHz, each in two polarizations, 
we obtained an average of 60 minutes of on-source 
per baseline under nominal observing conditions on 2004 February 27 (MJD 53062).  We 
used 0716+714 to calibrate phases and delays prior to self-calibration of the blazar,
using AIPS. We constructed deconvolved images with natural weighting of the data
(half-power beam width 1.55$\times$0.47mas at -2$^\circ$). The RMS background noise in
the image was  $\sim 10\%$ above the thermal noise limit of 0.18 mJy.  
The 7 mm observations recorded $\sim 41$~minutes on-source per baseline on
2004 March 3 (MJD 53067), tuned to 43.21 GHz.  About 30\% of the data could not 
be self-calibrated because of rapid ($\ll\,1$\,minute) tropospheric fluctuations.  
In addition, we obtained no usable data from the Los Alamos antenna, due 
to weather, and from one polarization-baseband combination recorded at 
the Saint Croix antenna.  The half-power beamwidth was $0.55\times0.30$~mas 
at $-3.7^\circ$ for natural weighting.

\begin{figure}[h!]
\plotone{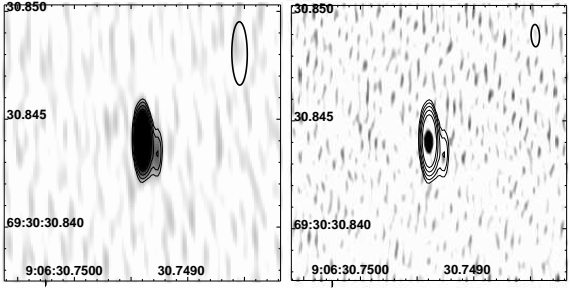}
\figcaption{\small
VLBA snapshot images of Q0906+6930. Left -- Emission at 2cm wavelength, with a 
peak flux of 115mJy/beam and contours at ($2^n$mJy/beam, n=0,1,..). A jet shows 
clearly at PA=225$^\circ$. Right -- Emission at 7mm (grayscale
with a hard stretch) and contours from the 2cm image overlayed. The jet
appears marginally detected at the $\sim3\sigma$ level.  The convolving beams
are shown (upper right) in each panel.
}
\end{figure}

	The compact core was unresolved at both wavelengths. In
the 2cm image a clear jet component is seen at PA 225$\pm 1^\circ$. For a two
component Gaussian model we fit a core flux of 115$\pm 0.3$mJy and a jet flux
of 6.3$\pm 0.4$mJy. The jet is marginally detected in our 7mm image with a 
maximum $\la 0.2$ beams from the position of the 2cm peak, although it is only 
comparable to the largest peaks in the image background. The two component
fit gives a core flux 42$\pm 1.9$mJy and jet flux $4.1\pm 1.1$mJy, for a 
3.7$\sigma$ detection.

Our combined {\it EGRET} likelihood analysis finds a nominal $\sim 1.5\sigma$ excess
of $\gamma$-ray photons at the radio position.  In the viewing period with 
the most
significant detection, VP0220, the source produced 1.12$\pm 0.76 \times 
10^{-7} {\gamma/ {\rm cm^2/s}}$\,(E$>$100\,MeV). However, if other sources
beyond the standard 4$\sigma$ 3EG sources are present in this region the
fit flux is even lower. At this point it is best to employ the mission-averaged
upper limit $4 \times 10^{-8} {\gamma/ {\rm cm^2/s}}$\,(E$>$100\,MeV).

\begin{figure}[h!]
\plotone{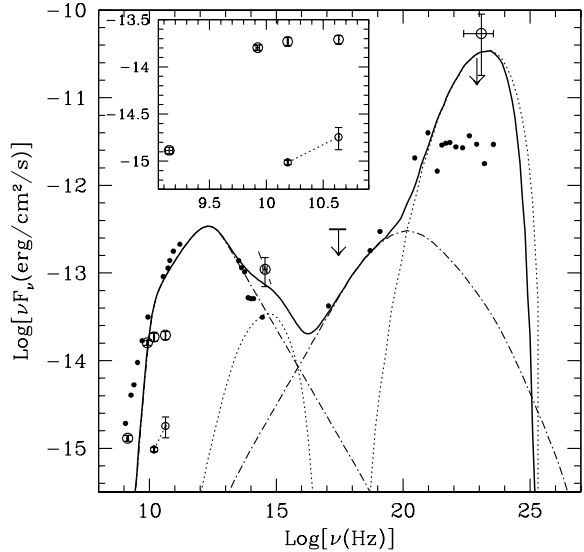}
\figcaption{\small Multiwavelength spectral energy distribution for Q0906+6930,
from non-simultaneous data (open circles).
The optical continuum flux is plotted with a line (short dash) extended to
show the slope. The X-ray upper limit is from the RASS.
For the {\it EGRET} band we show the brightest epoch flux and the mission-
averaged upper limit. For comparison we plot the SED of 3C 279 (points)
\citet{har01}, shifted to z=5.47.
A Comptonized spectrum (synchrotron self-Compton, dot-dash line; broad-line
region + Compton, dotted line) is plotted to guide the eye. The combined spectrum
(solid line) includes the effects of EBL absorption.
}
\end{figure}

	We can collect the existing data into a crude spectral
energy distribution (SED, Figure 3). Comparing to the
well known blazar 3C279 placed at z=5.47, Q0906+6930 has a brighter
UV flux, fainter cm wavelength emission and (potentially) a GeV peak
power nearly 10$\times$ greater. The source has by
far the highest radio loudness $R = f_{5GHz}/f_{0.44\mu,rest} \approx 10^3$
of any $z > 5$ QSO.  We also guide the eye with a simple synchrotron-Compton 
spectrum \citep{kra04} for a broken power-law approximating a cooled electron 
spectrum. The synchrotron component is broadly similar to that of 3C279.
The compact core spectrum (inset) appears to turn over at cm wavelengths and
does not connect  to the falling optical spectrum. This is not however
unexpected, as most blazars show extra radio components at cm wavelengths.
If the 7mm jet detection is trusted,
the $\alpha_{2cm/7mm} \approx 0.6$ of the jet component is somewhat flatter 
than that of the core $\alpha \approx 0.9$. This is unusual, although not
unique for pc-scale jets (A. Marscher, private comm.); additional VLBI 
observations will be needed to
see if this is temporary due to jet component brightening at 7mm.
With only a marginal GeV detection and an X-ray upper limit, the 
amplitude of the Compton component is poorly constrained. However, it
is interesting to note that the broad line region flux is 
much larger than that of 3C279 with with a $2-3\times$ brighter optical 
continuum and rest equivalent line widths $10\times$ larger. Since Comptonized
BLR flux is generally believed to dominate the GeV emission, a bright
$\gamma$-ray peak may be motivated, as shown by the dotted curves.

\section{Prospects for High Energy Detection}

	We expect the Compton scattered emission of blazar jets to be highly 
beamed. Thus, few blazar X- and $\gamma$-ray sources are visible, but these
may appear quite bright.  The GLAST
mission will have a large duty cycle and a survey
sensitivity $\sim 2 \times 10^{-9} \gamma/{\rm cm^2/s}
\approx 10^{-12} {\rm erg/cm^2/s}$\,(E$>$100\,MeV), so even if the source is a
flaring 3C279 analog, the prospects for detection are good. If the GeV spectrum of
this source can be measured, then GLAST can observe the effect of 
pair attenuation of the $>$GeV flux by the optical-IR EBL \citep{mp96}. 
At its high redshift $\le$10\,GeV observations of Q0906+6930 will be probing
$\le 3\mu$ photons from high $z$ star formation.  This EBL is
presently poorly understood, but an approximate attenuation curve based on
the lower $z$ EBL is applied for the combined spectrum (full line). The cut-off
is well within the GLAST range. It may be measured if, for example, a stable
absorption cutoff is seen on a GeV spectrum whose slope and amplitude varies
with time.

	In the X-ray regime, core emission probes the low energy 
Compton component, with the best connection to the radio. However,
most interesting is the possibility of detecting a resolved 
$\sim 1^{\prime\prime}$ ($\sim 10$kpc) scale X-ray jet produced by Compton
up-scatter of CMB photons. Such extended 
($\sim 2^{\prime\prime}$) jet structure has been seen in {\it CXO} 
observations of the QSO GB 1508+5714 \citep{sie03} at z=4.3.  This feature 
is also seen as a faint radio jet \citep{che04} and the flux ratio 
$f_X/f_R\sim 10^2$ 
supports the CMB up-scatter hypothesis.  Extended emission X-ray observed 
$\sim$20$^{\prime\prime}$ from the $z=5.99$ QSO J1306+0356 \citep{sch02b}
may represent a second high-z jet, although radio emission is not seen.
With a $(1+z)^4$ increase in the energy density of the CMB seed photons, such
jets may be detectable to very high redshifts\citep{sch02a}.  Q0906+6930 at z=5.47 
experiences $1752 \times$ the local CMB energy density.  As we already have
evidence for an energetic jet, prospects for detection of a Compton
up-scatter component are good and we are pursuing the required X-ray
and radio observations to test this hypothesis. If seen, the link to the
external seed field gives X-ray/radio measurements unique power to probe
the particle and field populations in this high-z jet.

\acknowledgments

{\small
This work was supported in part by NASA grants NAGS-13344 (RWR) and
NAS5-00147 (PFM) and SLAC/DOE contract DE-AC03-76SF00515. 

The Hobby-Eberly Telescope (HET) is a joint project of the University 
of Texas at Austin, the Pennsylvania State University, Stanford 
University, Ludwig-Maximilians-Universit\"at M\"unchen, and 
Georg-August-Universit\"at G\"ottingen. The HET is named in honor of 
its principal benefactors, William P. Hobby and Robert E. Eberly.
The Marcario Low Resolution Spectrograph is named for Mike Marcario 
of High Lonesome Optics who fabricated several optics for the instrument 
but died before its completion. The LRS is a joint project of the 
Hobby-Eberly Telescope partnership and the Instituto de Astronomia de 
la Universidad Nacional Autonoma de Mexico.
The National Radio Astronomy Observatory is a facility of the 
National Science Foundation operated under cooperative agreement
by Associated Universities, Inc.
}

\end{document}